\documentstyle[prb,aps]{revtex}
\def\rr{{{\bf r}}}
\def\rrp{{{\bf r^\prime}}}
\def\nup{{{n_\uparrow}}}
\def\ndo{{{n_\downarrow}}}

\begin{document}
%\draft

\title{ Volume change of bulk metals and metal clusters due to
spin-polarization}

\author{M. Payami }
\address{Center for Theoretical Physics and Mathematics,
Atomic Energy Organization of Iran,\\ P.~O.~Box 11365-8486, Tehran, Iran}

%\date{\today}
\maketitle
\begin{abstract}

The stabilized jellium model (SJM) provides us a method to calculate the volume
changes of different simple metals as a function of the spin polarization,
$\zeta$,
of the delocalized valence electrons. Our calculations show that for bulk
metals,
the equilibrium Wigner-Seitz (WS) radius, $\bar r_s(\zeta)$, is always an
increasing function of
the polarization i.e., the volume of a bulk metal always increases as $\zeta$
 increases, and the rate of increasing is
higher for higher electron density metals. Using the SJM along with the local
spin density approximation, we have also calculated the
equilibrium WS radius, $\bar r_s(N,\zeta)$, of spherical jellium clusters, at
which the pressure
on the cluster with given numbers of total electrons, $N$, and their spin
configuration $\zeta$ vanishes. Our calculations for Cs, Na, and Al clusters
show that $\bar r_s(N,\zeta)$ as a function of $\zeta$ behaves differently
depending on whether $N$ corresponds to a closed-shell or an open-shell
cluster. For a closed-shell cluster, it is an increasing function of $\zeta$
over the whole range $0\le\zeta\le 1$, whereas in open-shell clusters it has a
decreasing behavior over the range $0\le\zeta\le\zeta_0$, where $\zeta_0$ is a
polarization that the cluster has a configuration consistent with Hund's first
rule. The results show that for all neutral clusters with ground state spin
configuration, $\zeta_0$, the inequality $\bar r_s(N,\zeta_0)\le\bar
r_s(0)$ always holds (self-compression) but, at some polarization
$\zeta_1>\zeta_0$, the inequality changes the direction (self-expansion).
However, the inequality $\bar r_s(N,\zeta)\le\bar r_s(\zeta)$ always holds and
the equality is achieved in the limit $N\to\infty$.

\end{abstract}
%%%%%%%%%%%%%%%%%%%%%%%%%%%%%%%%%%%%%%%%%%%%%%%%%%%%%%%%%%%%%%%%%%%%%%%%%%%
%\newpage

\section{Introduction}
\label{sec1}

Jellium model (JM) with spherical geometry is the simplest model used in
theoretical study of simple metal clusters.\cite{ekardt,knight,brack93} In the
spherical JM, the
ions are smeared into a uniform positive charge background sphere of density
$n=3/4\pi r_s^3$ and radius $R=(zN)^{1/3}r_s$ where $z$ and $N$ are the valence
of the atom and the number of constituent atoms of the cluster, respectively.
In the simple JM calculations the
 value of $r_s$ is taken to be the bulk value of the Wigner-Seitz
(WS) radius of the metal. In some calculations involving metal surfaces,
one may use diffuse JM to take account of the ion relaxations at the surface
of that metal.\cite{rubio} However,
it is a well-known fact\cite{lang70,ashlang67} that the simple
JM yields negative surface energies at high electron densities ($r_s\le 2$),
and negative bulk moduli for $r_s\approx 6$.
Also, using the simple JM, it is not possible to find a realistic size and
energetics for metal clusters.
These drawbacks have roots in the mechanical instability\cite{shore99,pertran}
of the simple
JM system. That is, the bulk jellium system is stable only for $r_s=4.18$.
In 1990, Perdew {\it et al.} have introduced the stabilized
jellium\cite{pertran}
model (SJM) by adding two corrections to the simple JM energy. The first
correction subtracts the spurious self-energy of each WS cell from the JM
energy, and the second correction adds to the energy the effect of difference
in the
potential an electron sees from the discrete pseudoions and from the jellium
background.
The second correction introduces the core radius parameter, $r_c$, of the
pseudopotential which can be adjusted in such a way that the stability of the
bulk jellium system could be achieved at any observed $r_s$ value for the
valence electrons. Using the value $r_c^B$ for the core radius ( which
stabilizes the unpolarized bulk system) in the SJM energy functional of a
spin-polarized bulk system or a cluster, it is possible to find the equilibrium
$r_s$ value for any given spin polarization $\zeta$.

In this paper we have calculated the changes in the equilibrium $r_s$ values
due to polarization for bulk Cs, K, Na, Li, Ga, Al, and for Cs, Na, Al
clusters using the SJM in the framework
of local spin density approximation (LSDA). Our calculations show that for the
bulk system the volume is always an increasing function of the polarization and
the increasing rate is higher for higher electron density metals. Also, we have
solved
the self-consistent Kohn-Sham (KS) equations\cite{kohnsham} for Cs, Na, and Al
clusters
of different sizes in the spherical SJM and obtained the equilibrium sizes of
the clusters for each spin configuration consistent with Pauli's exclusion
principle. The results show that in a closed-shell cluster with $N$ total
electrons and polarization $\zeta$, the equilibrium WS radius
$\bar{r}_s(N,\zeta)$
 is an increasing function of $\zeta$
over the whole range $0\le\zeta\le 1$ whereas, for an
open-shell cluster (except for the two nearest neighbors of a closed-shell
cluster), it has a decreasing behavior over $0\le\zeta\le\zeta_0$, and an
increasing behavior over $\zeta_0\le\zeta\le 1$. Here, $\zeta_0$ corresponds to
an electronic spin configuration for which Hund's first rule is satisfied.
It has been shown previously\cite{pay99} that this property also holds in
the case of the simple JM with spherical geometry. Also, the equilibrium $r_s$
values for a neutral cluster are found
to be always smaller than that of the bulk metal with the same value of the
polarization $\zeta$. This effect is due to the surface tension which is
appreciable for small clusters. On the other hand, for an $N$ electron neutral
cluster there exists a polarization value $\zeta_1$, beyond which the
equilibrium $r_s$ value of the cluster exceeds the bulk $r_s$ value of
unpolarized metal. This is called self-expansion
\cite{perdew93,braj96,vieira97} and for two different
neutral clusters with the same number of electrons $N$, the $\zeta_1$ value of
the higher electron density metal is smaller than that of the lower density
metal.

The structure of this paper is as follows. In section \ref{sec2} we first show
the method by which one can calculate the changes in $r_s$ values due to spin
polarization for different bulk metals. We then explain how to apply the method
to metal clusters in order to calculate the equilibrium $r_s$ values for
different spin configurations. In section \ref{sec3} we present the results of
our calculations and finally, the work is concluded in section \ref{sec4}.
%%%%%%%%%%%%%%%%%%%%%%%%%%%%%%%%%%%%%%%%%%%%%%%%%%%%%%%%%%%%%%%%%%%%%%%%%%%

\section{Calculational Scheme}
\label{sec2}

In the context of the SJM, the average energy per valence electron in the
bulk
with density parameter $r_s$ and polarization $\zeta$ is given by\cite{pay98}

\begin{equation}
\varepsilon(r_s,\zeta,r_c)=t_s(r_s,\zeta)+\varepsilon_{xc}(r_s,\zeta)+\bar
w_R(r_s,r_c)+\varepsilon_{\rm M}(r_s),
\label{eq1}
\end{equation}

where

\begin{equation}
t_s(r_s,\zeta)=\frac{c_k}{r_s^2} \left[(1+\zeta)^{5/3}+(1-\zeta)^{5/3}\right]
\label{eq2}
\end{equation}

\begin{equation}
\varepsilon_{xc}(r_s,\zeta)=\frac{c_x}{r_s}
\left[(1+\zeta)^{4/3}+(1-\zeta)^{4/3}\right]+\varepsilon_c(r_s,\zeta)
\label{eq3}
\end{equation}

\begin{equation}
c_k=\frac{3}{10} \left(\frac{9\pi}{4}\right)^{2/3}
;\;\;\;\;c_x=\frac{3}{4}\left(\frac{9}{4\pi^2}\right)^{1/3}.
\label{eq4}
\end{equation}

All equations throughout this paper are expressed in Rydberg atomic units. Here
$t_s$ and $\varepsilon_{xc}$ are the mean noninteracting kinetic energy and
the exchange-correlation
energy per particle, respectively. For $\varepsilon_c$ we use the Perdew-Wang
parametrization. \cite{perwan}
Here, $\varepsilon_{\rm M}$ is the average Madelung energy,
$\varepsilon_{\rm M}=-9z/5r_0$, and $r_0$ is the radius of the WS sphere,
$r_0=z^{1/3}r_s$. We set for monovalent metals $z=1$, and for polyvalent
metals
we set $z^*=1$ (for details see Ref.[\ref{pertran}]).
In Eq.(\ref{eq1}), $\zeta=(n_\uparrow-n_\downarrow)/(n_\uparrow+n_\downarrow)$
in which $n_\uparrow$ and $n_\downarrow$ are the spin densities of the
homogeneous system with total
density $n=n_\uparrow+n_\downarrow$. The quantity $\bar w_R$ is the
average value (over the WS cell) of the repulsive part of the Ashcroft empty
core\cite{ash66} pseudopotential,

\begin{equation}
w(r)=-\frac{2z}{r}+w_R,\;\;\;\;\;w_R=+\frac{2z}{r}\theta(r_c-r),
\label{eq5}
\end{equation}
and is given by
$\bar
w_R=3r_c^2/r_s^3$ where, $z$ is the valence of the atom,
$\theta(x)$ is the ordinary step function which assumes the value of unity for
positive arguments, and zero for negative values.

The core radius is fixed to the bulk value, $r_c^B$,
 by setting the pressure of the unpolarized bulk system equal to zero
at the observed equilibrium density
$\bar{n}=3/4\pi\bar{r}_s^3$:

\begin{equation}
\left.\frac{\partial}{\partial r_s}\varepsilon(r_s,0,r_c)\right|_{
 r_s=\bar r_s,r_c=r_c^B}=0.
\label{eq6}
\end{equation}
The derivative is taken at fixed $r_c$, and the solution of the above
equation gives $r_c^B$ as a function of
$\bar{r}_s$

\begin{equation}
r_c^B(\bar r_s)=\frac{\bar r_s^{3/2}}{3}\left\{\left[
-2t_s(r_s,0)-\varepsilon_x(r_s,0)+ r_s
\frac{\partial}{\partial
 r_s}\varepsilon_c(r_s,0)-\varepsilon_M(
 r_s)\right]_{r_s=\bar r_s}\right\}^{1/2}.
\label{eq7}
\end{equation}
Here, $\bar r_s\equiv\bar r_s(\zeta=0)$
is the observed equilibrium density parameter for the unpolarized bulk system,
and takes the values of 2.07, 2.19, 3.28, 3.99, 4.96, 5.63 for Al, Ga, Li, Na,
K, Cs, respectively.
Inserting the $r_c^B$ from Eq. (\ref{eq7}) into Eq. (\ref{eq1}), the
equilibrium $r_s$ value of the polarized bulk system is obtained by the
solution of the equation

\begin{equation}
\left.\frac{\partial}{\partial r_s}\varepsilon(r_s,\zeta,r_c^B)\right|_{
 r_s=\bar r_s}=0.
\label{eq8}
\end{equation}
The derivative is taken at fixed $\zeta$ and $r_c^B$. The solution gives
the equilibrium density parameter $\bar r_s(\zeta)$ as a function of the
polarization. In this procedure, by taking a constant value for the core radius
of the pseudopotential, we have assumed that the core region of an
atom is rigid and does not change in the process of the spin polarization of
the delocalized valence electrons. This approximation works well when the
distance between the neighboring atoms is sufficiently larger than the
extension of the core electron orbital wave functions.

The SJM energy for a spin-polarized system with boundary surface is
\cite{pertran}

\begin{eqnarray}
E_{\rm SJM}\left[\nup,\ndo,n_+\right]&=&
E_{\rm JM}\left[\nup,\ndo,n_+\right]+\left(\varepsilon_M(r_s)+\bar
w_R(r_s,r_c^B)\right)\int d\rr\;n_+(\rr) \nonumber \\
  &&+\langle\delta v\rangle_{\rm WS}(r_s,r_c^B)\int
d\rr\;\Theta(\rr)\left[n(\rr)-n_+( \rr)\right],
\label{eq9}
\end{eqnarray}
where
\begin{eqnarray}
E_{\rm
JM}\left[\nup,\ndo,n_+\right]&=&T_s\left[\nup,\ndo\right]+E_{xc}\left[\nup,\ndo
\right] \nonumber\\ &&+\frac{1}{2}\int
d\rr\;\phi\left([n,n_+];\rr\right)\left[n(\rr)-n_+(\rr)\right]
\label{eq10}
\end{eqnarray}
and
\begin{equation}
\phi\left([n,n_+];\rr\right)=2\int
d\rrp\;\frac{\left[n(\rrp)-n_+(\rrp)\right]}{\left|\rr-\rrp\right|}.
\label{eq11}
\end{equation}
Here, $n=n_\uparrow+n_\downarrow$ and $\Theta(\rr)$ takes the value of unity
inside the jellium background and zero, outside. The first and second terms in
the right hand side of Eq.(\ref{eq10}) are
the non-interacting kinetic energy and the exchange-correlation energy, and the
last term is the Coulomb interaction energy of the system.
The effective potential, used in the self-consistent KS equations, is obtained
by
taking the variational derivative of the SJM energy functional with respect to
the spin densities as

\begin{eqnarray}
v_{eff}^\sigma\left(\left[n_\uparrow,n_\downarrow,n_+\right];\rr\right)&=&
\frac{\delta} {\delta n_\sigma(\rr)}(E_{\rm SJM} -T_s)\nonumber\\
&=&\phi\left(\left[n,n_+\right];\rr\right)+
v_{xc}^\sigma\left(\left[n_\uparrow,n_\downarrow\right];\rr\right)
 +\Theta(\rr)\langle\delta v\rangle_{\rm WS} (r_s,r_c^B),
\label{eq12}
\end{eqnarray}
where $\sigma=\uparrow,\downarrow$.
By solving the KS equations
\begin{equation}
\left(\nabla^2+v_{eff}^\sigma(\rr)\right)\phi_i^\sigma(
\rr)=\varepsilon_i^\sigma
 \phi_i^\sigma(\rr),\;\;\;\;\;\;\;\sigma=\uparrow,\downarrow,
\label{eq13}
\end{equation}

\begin{equation}
n(\rr)=\sum_{\sigma=\uparrow,\downarrow}n_\sigma(\rr),
\label{eq14}
\end{equation}

\begin{equation}
n_\sigma(\rr)=\sum_{i(occ)}\left|\phi_i^\sigma(\rr)\right|^2,
\label{eq15}
\end{equation}
and finding the self-consistent values for $\varepsilon_i^\sigma$ and
$\phi_i^\sigma$, one obtains the total energy.

In our spherical JM, we have

\begin{equation}
n_+(\rr)=\frac{3}{4\pi r_s^3}\theta(R-r)
\label{eq16}
\end{equation}
in which $R=(zN)^{1/3}r_s$ is the radius of the jellium sphere, and $n(\rr)$
denotes the electron density at point $\rr$ in space.
The quantity $\langle\delta v\rangle_{\rm WS}$
is the average of the difference potential over the
Wigner-Seitz cell and the difference potential, $\delta v$, is defined as the
difference between the pseudopotential of a lattice of ions and the
electrostatic potential of the jellium positive background.
Using the Eq. (21) of Ref. [\ref{pertran}], this average value is given by

\begin{equation}
\langle\delta v\rangle_{\rm WS}(r_s,r_c^B)=\frac{3(r_c^B)^2}{r_s^3}-
\frac{3}{5r_s}.
\label{eq17}
\end{equation}

Applying Eq. (\ref{eq9}) to a metal cluster which contains $N_\uparrow$
spin-up,
 $N_\downarrow$ spin-down and $N$ $(=N_\uparrow+N_\downarrow)$ total electrons,
the
 SJM energy becomes a function of $N$,
$\zeta\equiv(N_\uparrow-N_\downarrow)/N$,
 $r_s$, and $r_c^B$. The equilibrium density parameter, $\bar r_s(N,\zeta)$,
for a cluster is the solution of the equation

\begin{equation}
\left.\frac{\partial}{\partial r_s}E(N,\zeta,r_s,r_c^B)\right|_{r_s=\bar
r_s(N,\zeta)}=0.
\label{eq18}
\end{equation}
Here again, the derivative is taken at fixed values of $N$, $\zeta$, and
$r_c^B$.
For an $N$-electron
cluster, we have solved the KS equations\cite{kohnsham}
self-consistently for various spin configurations and $r_s$ values and obtained
the equilibrium density
parameter, $\bar r_s(N,\zeta)$, and its corresponding energy,
$\bar E(N,\zeta)\equiv E(N,\zeta,\bar r_s(N,\zeta),r_c^B)$
for each allowed spin configuration.

%%%%%%%%%%%%%%%%%%%%%%%%%%%%%%%%%%%%%%%%%%%%%%%%%%%%%%%%%%%%%%%%%%%%%%%

\section{Results}
\label{sec3}
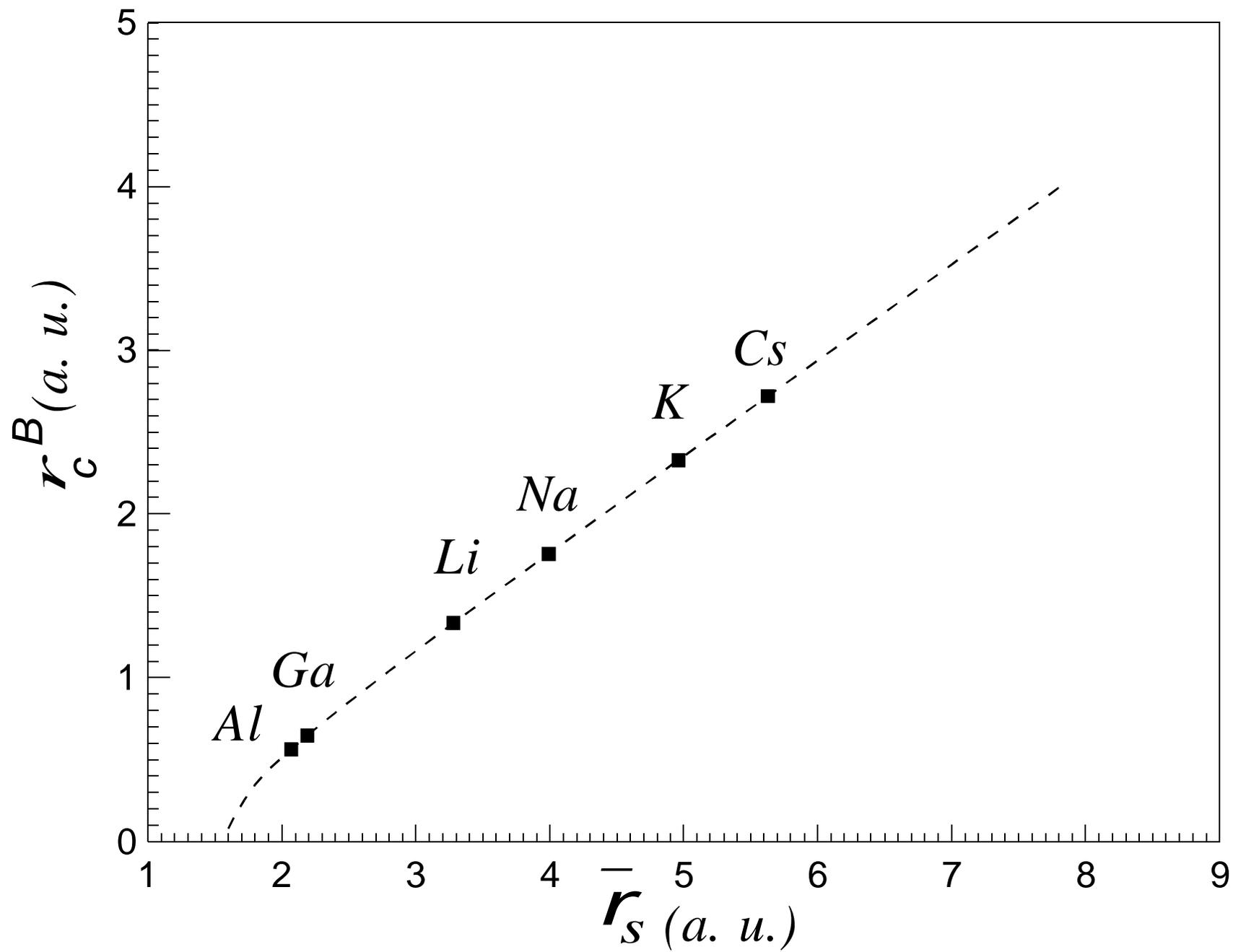
\begin{figure}
\caption{The pseudopotential core radius, $r_c^B$, in atomic units as
a function of the equilibrium $r_s$ value of the bulk metal. The rigid squares
specify different metals. The plot shows a linear behavior at rather low
densities.}
\label{fig1}
\end{figure}

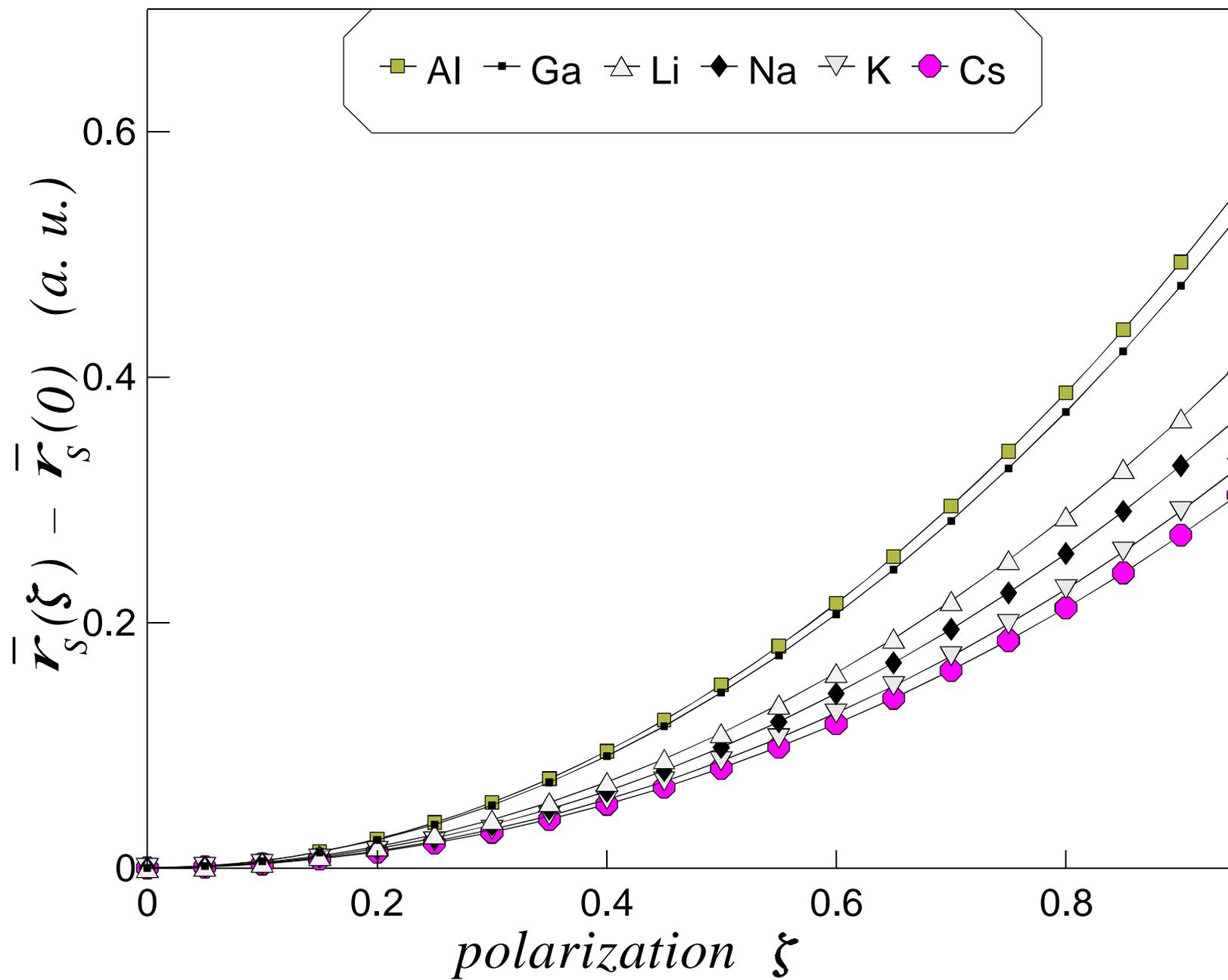
\begin{figure}
\caption{The changes in the equilibrium WS radius relative to the
unpolarized value, [$\bar r_s(\zeta)-\bar r_s(0)$], in atomic units, for
different metals as functions of the polarization, $\zeta$. All have increasing
behaviors and the increasing rate is higher for higher electron density metals.
 }
\label{fig2}
\end{figure}

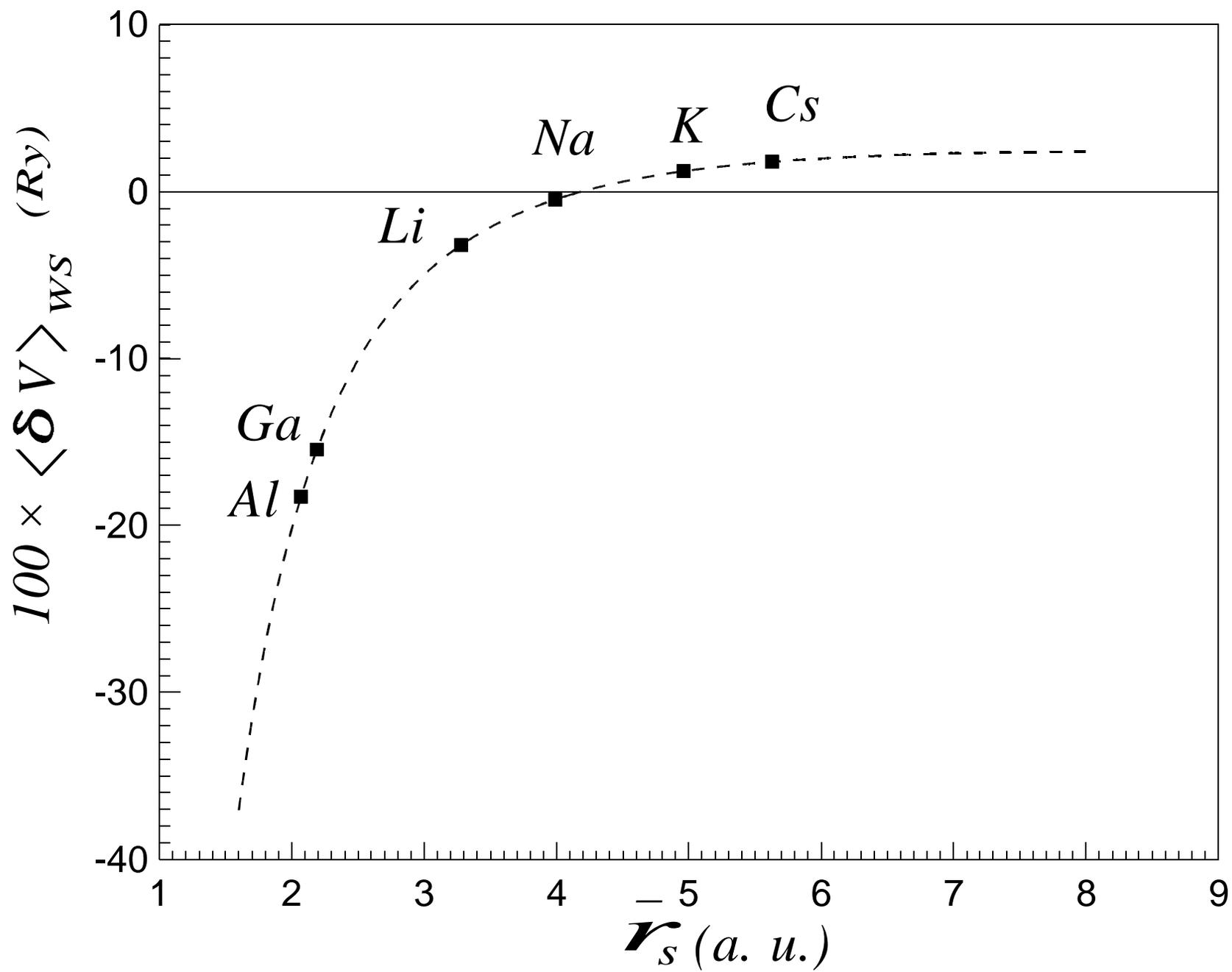
\begin{figure}
\caption{The average value of the difference potential, in Rydbergs, as a
function of the equilibrium WS radius for a bulk jellium system. The rigid
squares correspond to different metals.}
\label{fig3}
\end{figure}

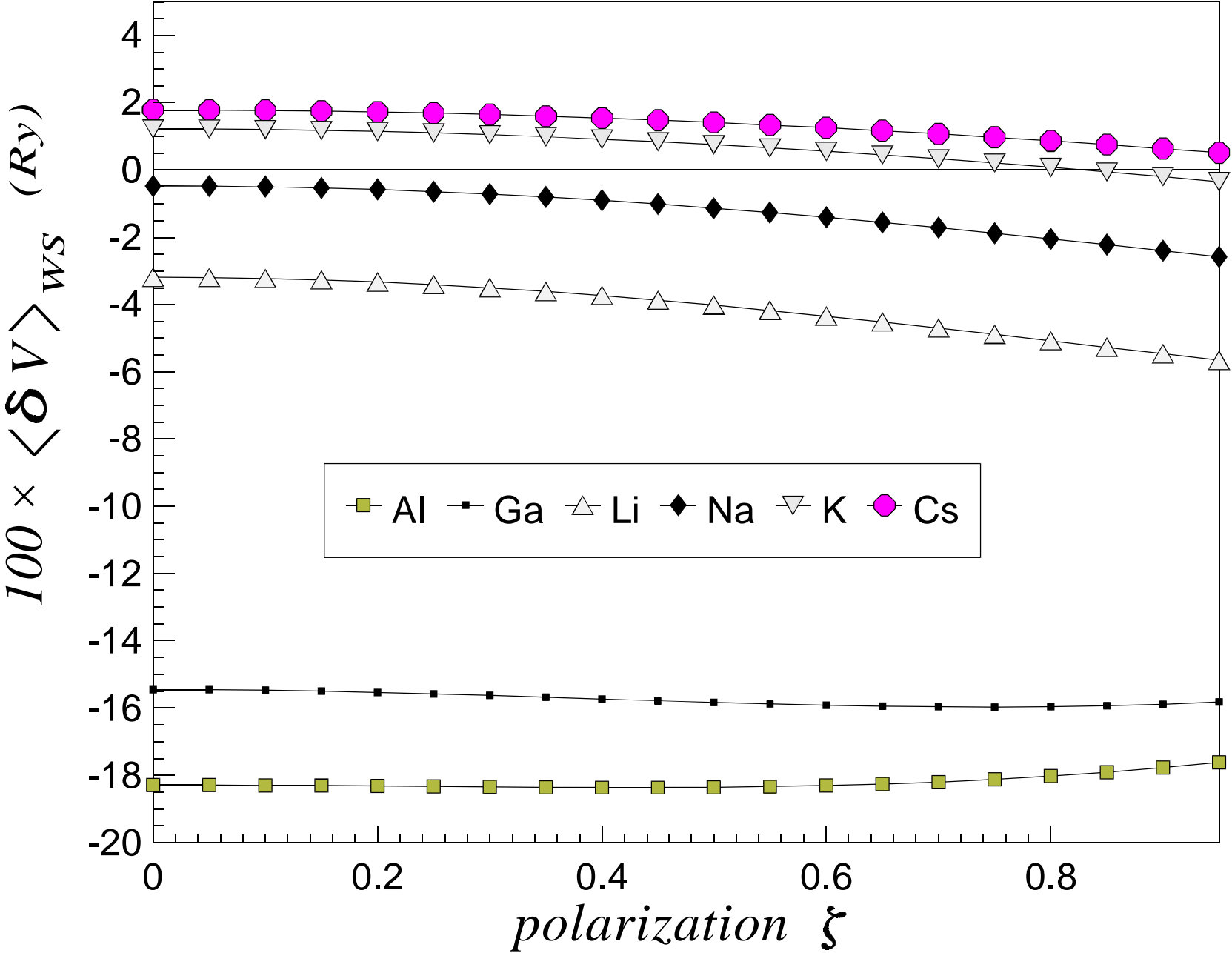
\begin{figure}
\caption{The average values of the difference potential, in Rydbergs, as
functions of the polarization for different bulk metals. The plots show
decreasing behaviors for Cs, K, Na, Li over the whole range; whereas the plots
for Ga and Al show decreasing behaviors at low polarizations, and increasing
behaviors at higher polarizations with minima in between.}
\label{fig4}
\end{figure}

In Fig. \ref{fig1} we have shown the $r_c^B$ values obtained from the
application of Eq.
(\ref{eq7}) to different $r_s$ values. Different metals are specified by the
rigid squares. It is seen that the plot shows a linear behavior for relatively
low electron density metals. Inserting these values of $r_c^B$ into Eq.
(\ref{eq1}) and solving the Eq. (\ref{eq8}) for different values of
polarization, $\zeta$, we have obtained the equilibrium $r_s$ values for
different bulk metals at various polarizations. The values obtained at
$\zeta=1$ are 2.62, 2.72, 3.69, 4.36, 5.28, and 5.93 for Al, Ga, Li, Na, K, and
Cs, respectively. The result for all polarizations $(0\le\zeta\le 1)$ is shown
in
Fig. \ref{fig2}. In this figure, we have plotted the changes in the equilibrium
$r_s$ values relative to the unpolarized value as a function of the
polarization for different metals. The results show an increasing behavior for
all metals but, the increasing rate is higher for higher electron density
metals.
This property can be easily explained in terms of the Fermi holes arround the
electrons seen by other electrons with the same quantum numbers. In the case of
very low electron density metals, the average distance between the electrons is
much more larger than the effective range of the Fermi hole. Therefore, if some
or all of the spin-down electrons undergo a spin flip and change to spin-up
electrons, the number of spin-up electron Fermi holes will increase. But,
these Fermi holes will not have any overlap and therefore,
effectively no repulsive force will act on the electrons to increase the volume
of the system. On the other hand, if the average distance between the electrons
in a metal be much less than the range of a Fermi hole, then these spin flips
cause the Fermi holes to overlap and as a result a large repulsive force will
act on the electrons to increase the volume of the system.
This explains the higher increasing rate for Al and the lower increasing rate
for Cs metals.
Figure \ref{fig3} shows the plot of
$\langle\delta v\rangle_{\rm WS}\left(\bar r_s,r_c^B(\bar r_s)\right)$,
obtained from Eq. (\ref{eq17}) and Eq. (\ref{eq7}), as a function of the bulk
 equilibrium WS radius, $\bar r_s$. The rigid squares correspond to
different metals. The value of $\langle\delta v\rangle_{\rm WS}$ for Na is
vanishingly small because, the
equilibrium $r_s$ value of sodium is very close to 4.18 at which this average
vanishes. For values of $r_s$ greater than 4.18, the correction to the KS
effective potantial is positive [see Eq.(\ref{eq12})] and this decreases the
well depth and cause the electrons relax outward to make the pressure on the
system (due to jellium model) to vanish. In this case, the leakage of the
electrons across the jellium boundary surface is increased relative to the
simple JM case. On the other hand, for
$r_s<4.18$ the potential depth is increased by the correction and the electrons
should decrease their relative mean distance to make the pressure on the
jellium system to vanish. Therefore, the leakage of the electrons decreases
relative to the
simple jellium model case. In Fig. \ref{fig4},
using the Eqs. (\ref{eq7}), (\ref{eq8}), and (\ref{eq17} )
we
have plotted the variation of
 $\langle\delta v\rangle_{\rm WS}\left(\bar r_s(\zeta),r_c^B)\right)$
 as a function
of $\zeta$ for different metallic densities.
For low electron density metals, it has a decreasing behavior and in the case
of K metal, there is a change in the sign at high polarizations. However, for
Al and Ga, it shows rather different behaviors. That is, all Cs, K, Na, Li show
a decreasing behavior over the whole range $0\le\zeta\le 1$ whereas, Al and Ga
have decreasing behaviors at lower polarizations and increasing behaviors at
higher polarizations. These different behaviors can be explained by the
detail analysis of the Eq.(\ref{eq17}). The right hand side of Eq.(\ref{eq17})
has a physical minimum at $r_s=\sqrt{15}r_c^B$. In order to realize this
minimum for a metal over the range $0\le\zeta\le 1$, the core radius of that
metal should satisfy the inequality

\begin{equation}
\frac{\bar r_s(0)}{\sqrt{15}}\le r_c^B\le\frac{\bar r_s(1)}{\sqrt{15}}.
\label{eq18-1}
\end{equation}
In the above inequatity, $\bar r_s(0)$ and $\bar r_s(1)$ are the equilibrium
values at $\zeta=0$ and $\zeta=1$, respectively.
Examining for different metals show that, out of the six
mentioned metals, only Al and Ga with respective core radius values of 0.56 and
0.65 satisfy this constraint.
Comparing this figure with Fig. 1 of Ref. [\ref{pay98}],
reveals the different behaviors for high electron density metals, predicted by
the stabilized spin-polarized
jellium model\cite{pay98} (SSPJM). The different behavior in the SSPJM has
roots in the rough estimation used there for the increment in $r_s$ due to
polarization. But here, we could calculate the increment for each metal
separately, as shown in Fig. \ref{fig2}.

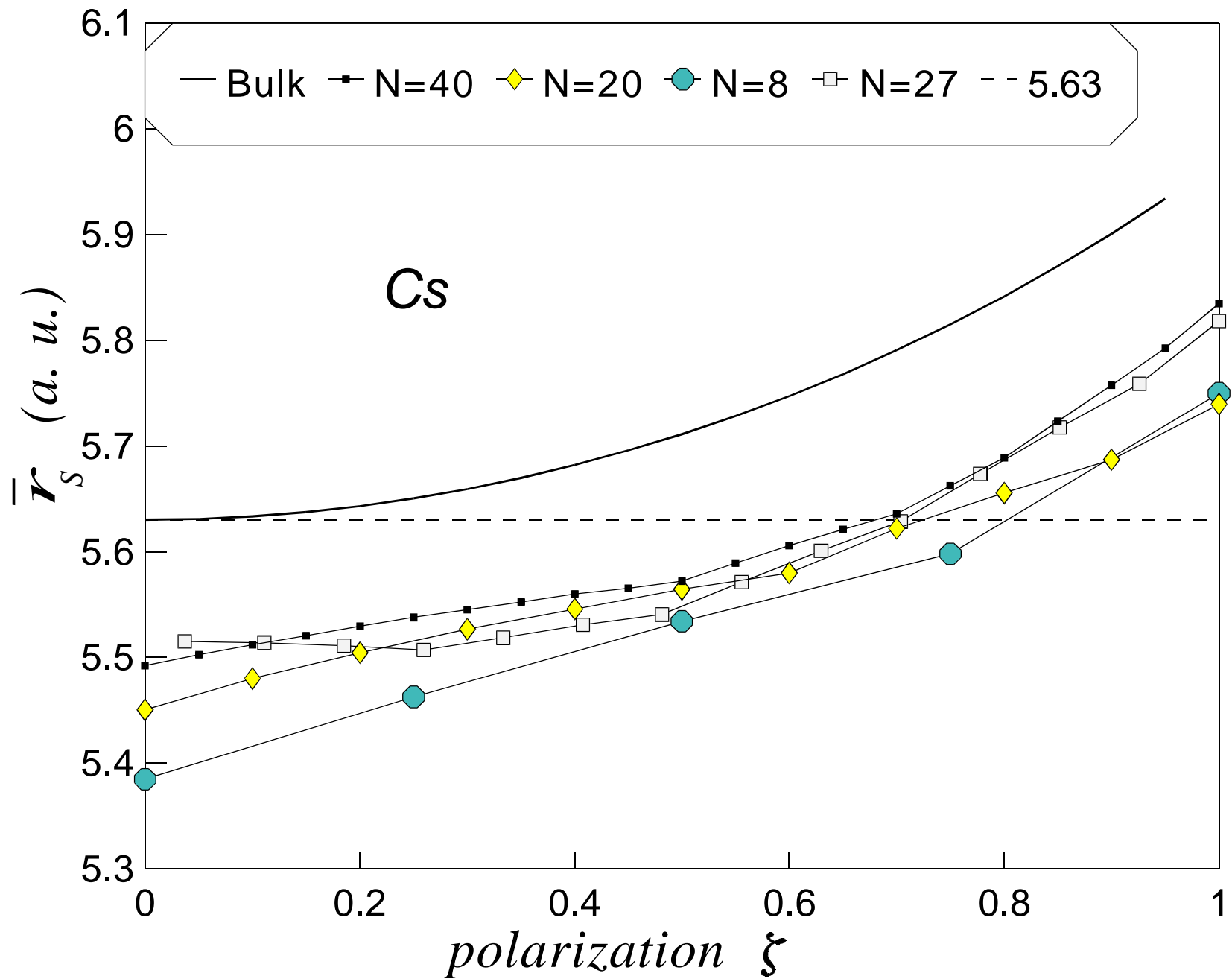
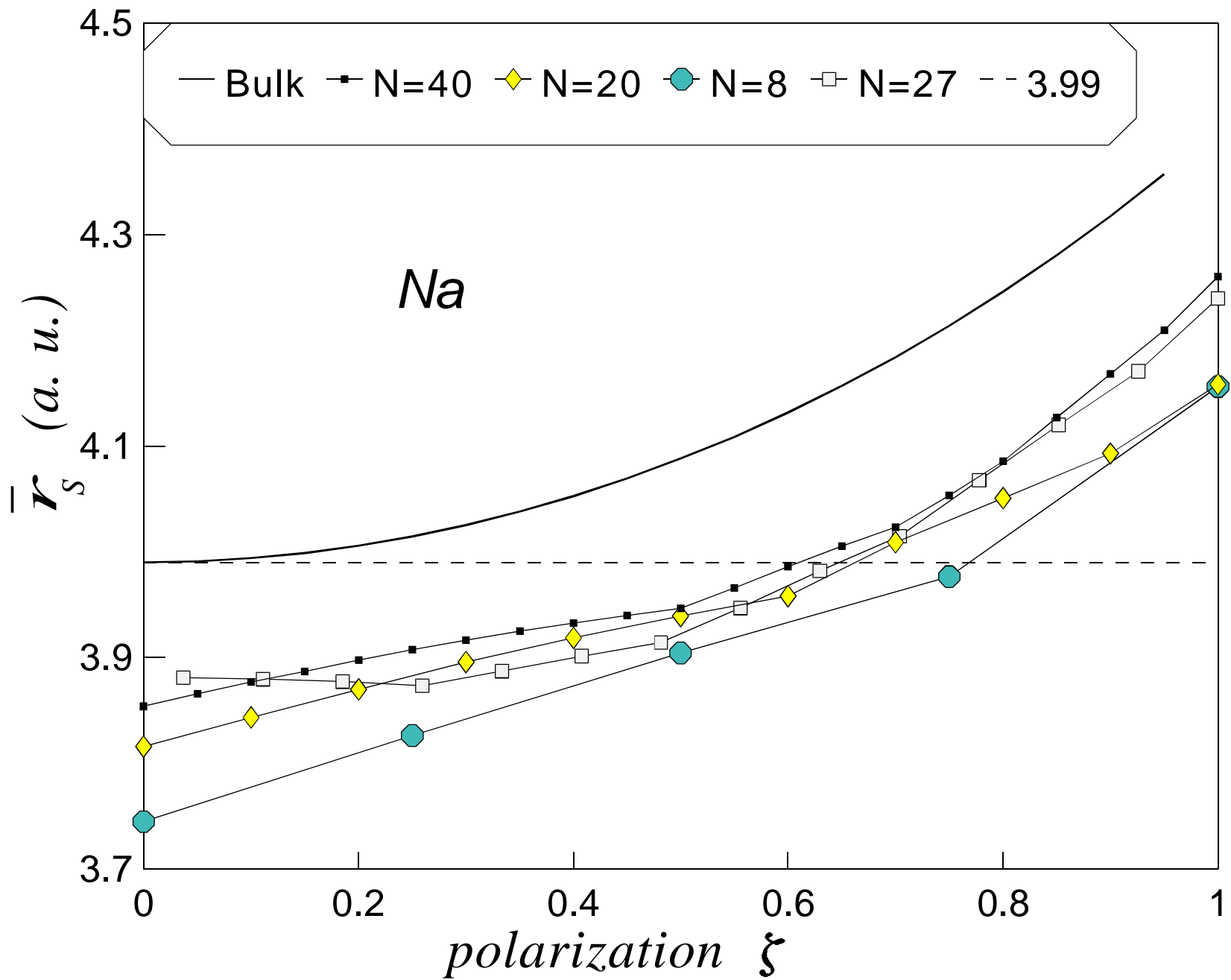
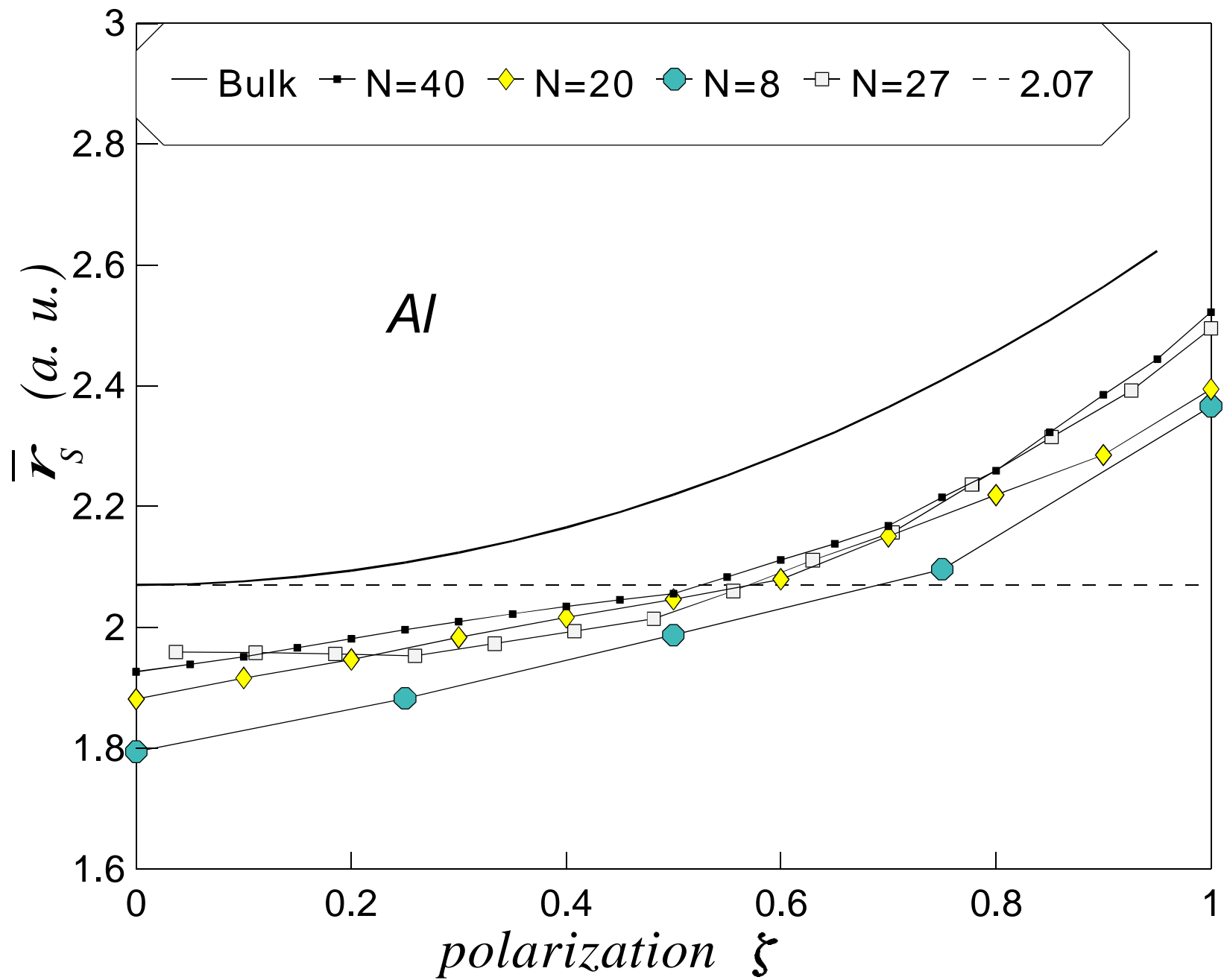
\begin{figure}
\caption{The equilibrium WS radius, $\bar r_s(N,\zeta)$, in atomic units, as
functions of the polarization for (a)- Cs, (b)- Na, (c)- Al clusters.
In all closed-shell clusters (here, $N$=8,20,40) $\bar r_s(N,\zeta)$ is an
increasing function of $\zeta$ while for the open-shell cluster (here, $N=27$)
it is decreasing over $0\le\zeta\le 7/27$, and increasing over $7/27\le\zeta\le
1$. The solid lines correspond to the bulk $\bar r_s(\zeta)$, and the dashed
lines correspond to the equilibrium value of the unpolarized bulk system, $\bar
r_s(0)$.}
\label{fig5}
\end{figure}

In order to find the equilibrium size of an $N$ electron cluster in the SJM for
various spin
configurations of $N_\uparrow-N_\downarrow=0$, $N_\uparrow-N_\downarrow=2$,
$N_\uparrow-N_\downarrow=4$, $\cdots$, $N_\uparrow-N_\downarrow=N_\uparrow$,
keeping the total number of electrons, $N_\uparrow+N_\downarrow=N$, fixed;
we solve the Eq. (\ref{eq18}) using the set of self-consistent Eqs.
(\ref{eq12})-(\ref{eq15}). In the above we have assumed an even number of
total electrons, $N$. For an odd number of electrons, the differences
($N_\uparrow-N_\downarrow$) would also be odd numbers.
In Figs. \ref{fig5}(a)-(c) we have plotted the equilibrium $r_s$ values,
$\bar r_s(N,\zeta)$,
obtained for different spin configurations,  as a function
of $\zeta\equiv(N_\uparrow-N_\downarrow)/N$ for Cs, Na, and Al clusters.
To clarify the different behaviors of the closed-shell and open-shell clusters,
we have studied the $N$=8, 20, 40 cases which are closed-shell clusters and
$N$=27 which is an open-shell cluster. We have also compared the result with
the bulk case.
In Figs. \ref{fig5}, the dashed lines correspond to the equilibrium WS radius
of the unpolarized bulk metal, $\bar r_s(\zeta=0)$. In all the closed-shell
clusters, $\bar r_s(N,\zeta)$ is an increasing function of $\zeta$ while for
the open-shell cluster, $N=27$, it is decreasing over
$0\le\zeta\le 7/27$,
and increasing over $7/27\le\zeta\le 1$. Here, $\zeta_0=7/27$ corresponds to an
electronic spin configuration for which Hund's first rule is satisfied.
That is, the configuration in which the up-spin shell with $l=3$ is half
filled. This difference between closed-shell and open-shell clusters can be
explained as follows. For an open-shell cluster if one increases the spin
polarization from the possible minimum value consistent with the Pauli
exclusion principle, one should make a spin-flip in the last uncomplete shell.
Because of high degeneracy for the spherical geometry, this spin-flip in the
last shell does not change appreciably the kinetic energy contribution to the
total energy but changes appreciably the exchange-correlation energy which in
turn gives rise to a deeper effective potential that makes the KS orbitals more
localized and therefore a smaller size for the cluster.
On the other hand, in closed-shell clusters increasing the polarization is
always accompanied by transition of electrons to unoccupied shells that have
larger kinetic energies which leads to larger cluster sizes.
In Figs. \ref{fig5} we notice that as the size of the cluster increases, the
plot of $\bar r_s(N,\zeta)$ resembles much more to the bulk function
$\bar r_s(\zeta)$ and approaches to it so that, as is expected, we have

\begin{equation}
\lim_{N\to\infty}\bar r_s(N,\zeta)=\bar r_s(\zeta).
\label{eq19}
\end{equation}

As we see, in all neutral clusters, the inequality
\begin{equation}
\bar r_s(N,\zeta)\le\bar r_s(0)
\label{eq20}
\end{equation}
always holds for the ground state of the cluster in which $\zeta=\zeta_0$. This
effect is called
self-compression\cite{perdew93} and is due to surface tension. Now if we
increase the polarization of
the cluster, $\zeta\equiv(N_\uparrow-N_\downarrow)/N$ relative to $\zeta_0$ and
obtain the equilibrium WS radius by solving the Eq.(\ref{eq18}), we see that
beyond some polarization, $\zeta_1$, the inequality in Eq.(\ref{eq20}) changes
the direction and the equilibrium $r_s$ value of the cluster exceeds the bulk
value $\bar r_s(0)$. This is called self-expansion which is also observed in
charged metal clusters.\cite{braj96} This value of $\zeta_1$ is shifted toward
zero as the size of the cluster increases. Also, comparison of Figs.
\ref{fig5}(a)-(c) show that for two clusters with the same $N$, the value of
$\zeta_1$ is smaller for higher electron density metal.

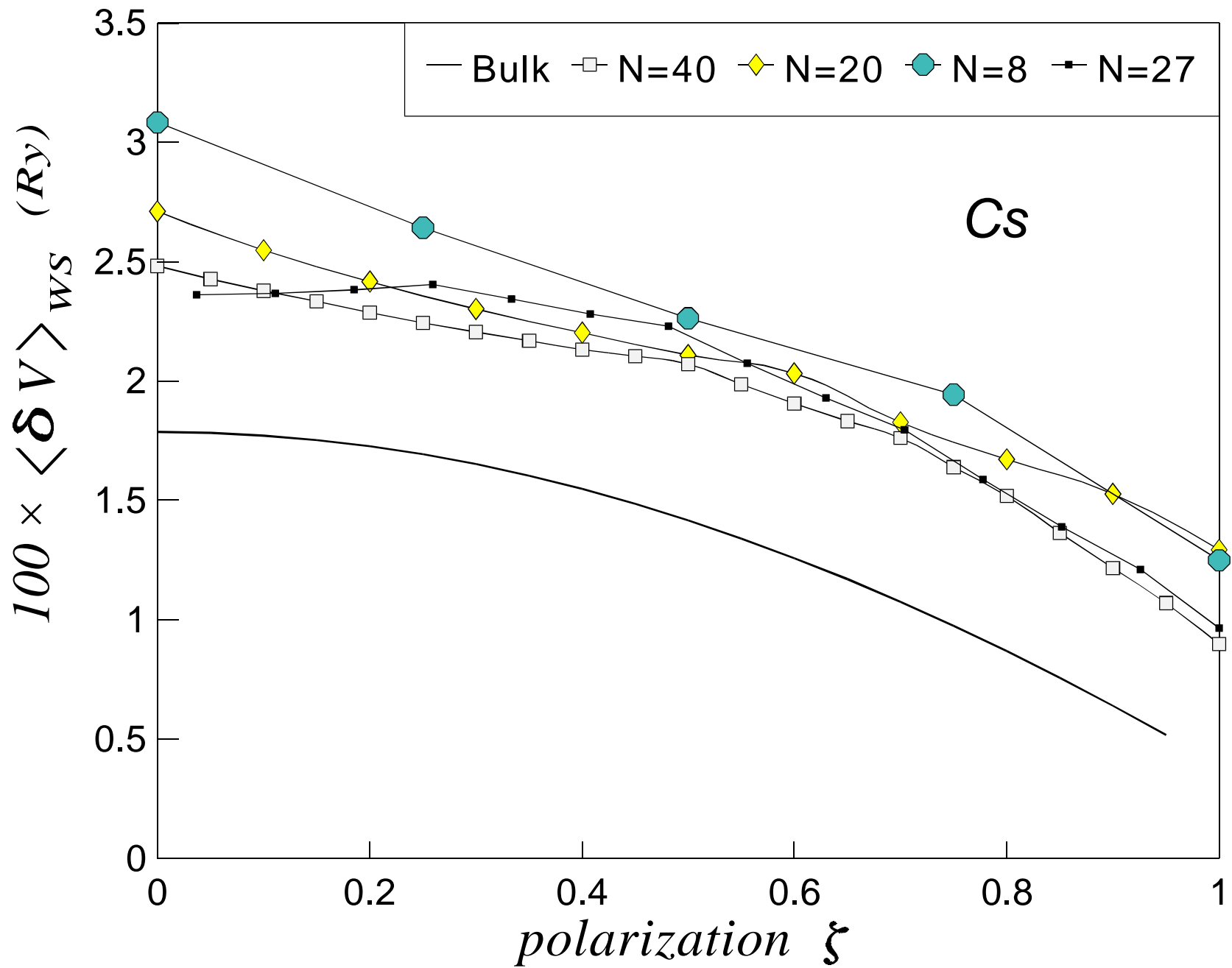
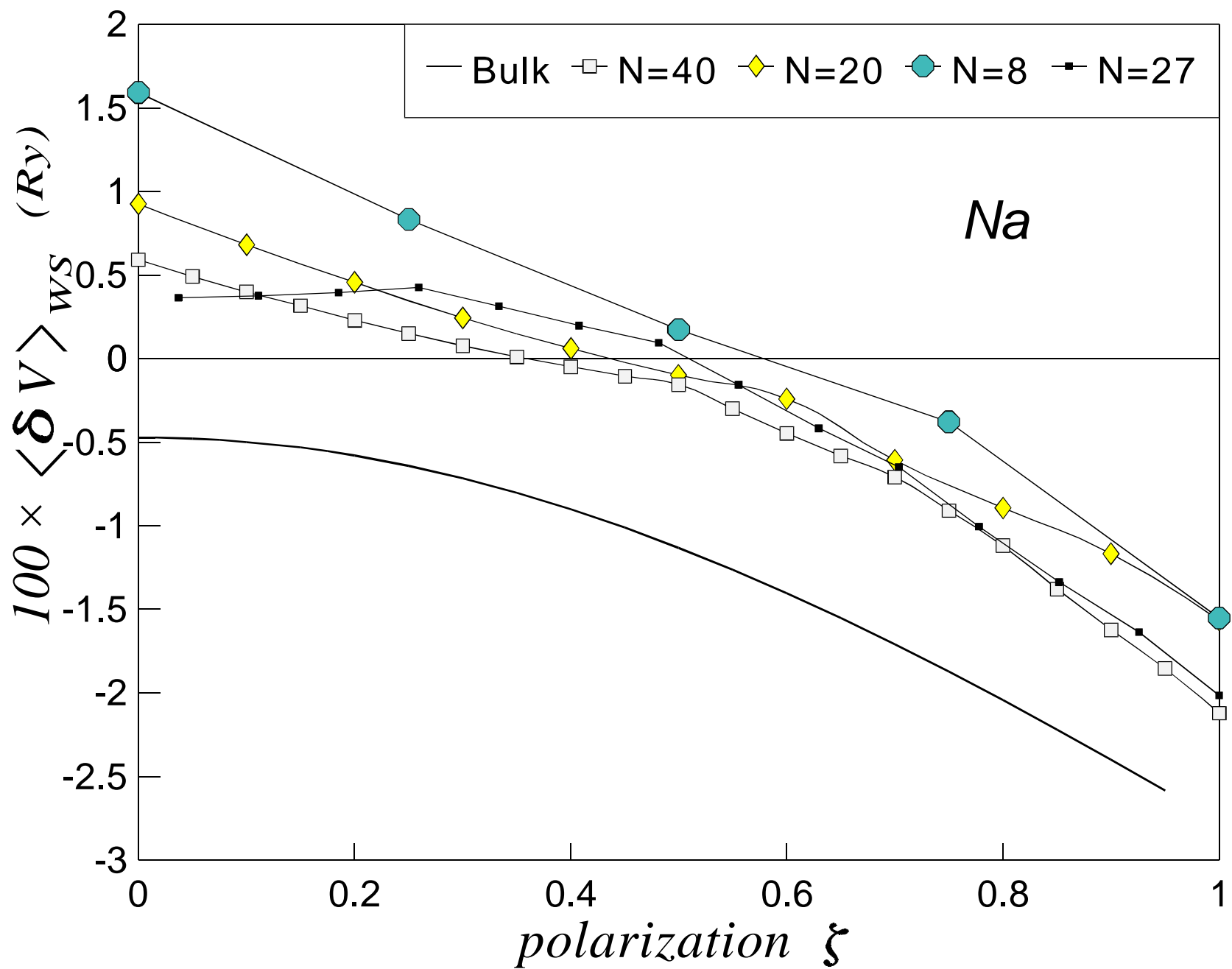
\begin{figure}
\caption{The average values of the difference potential, in Rydbergs, as
functions of the spin configurations, $\zeta$. In (a)- Cs and (b)- Na,
for the closed-shell clusters (here, $N$=8,20,40)  it
shows a decreasing behavior over the whole range, the same as the bulk case
which is specified by a solid line. But, in open-shell clusters (here,
$N=27$) it is increasing over $0\le\zeta\le 7/27$, and decreasing over
$7/27\le\zeta\le 1$. In (c)- Al, the same behaviors hold for lower
polarizations but, as discussed in the text, it shows a minimum at higher
polarizations. }
\label{fig6}
\end{figure}

In Figs. \ref{fig6}(a)-(c), we have plotted the quantity
$\langle\delta v\rangle_{\rm WS}$ for Cs, Na, Al clusters of different sizes,
as a function of $\zeta$, using the equilibrium values $\bar r_s(N,\zeta)$ in
the Eq.(\ref{eq17}); and have compared with their respective bulk functions. It
is seen that for metals which the inequality in Eq.(\ref{eq19}) does not hold,
this quantity has a decreasing behavior whenever $\bar r_s(N,\zeta)$ has an
increasing behavior, and vice versa. For Cs clusters in Fig.\ref{fig6}(a), the
quantity has positive values over the whole range $0\le\zeta\le 1$. In Fig.
\ref{fig6}(b), for Na clusters, this quantity changes sign beyond some $\zeta$
value. The Al clusters in Fig. \ref{fig6}(c) have negative values for
$\langle\delta v\rangle_{\rm WS}$ and they show minima as discussed before.

%%%%%%%%%%%%%%%%%%%%%%%%%%%%%%%%%%%%%%%%%%%%%%%%%%%%%%%%%%%%%%%%%%%%%

\section{Summary and Conclusion}
\label{sec4}
In this work, we have calculated the equilibrium $r_s$ values of different
metals as a function of their electronic spin polarizations, using the
stabilized
jellium model along with the local spin density approximation. Our calculations
show an increasing behavior for the bulk Wigner-Seitz radius of electron as a
function of polarization. Also we have shown that the increasing rate is higher
for higher electron density metals. Calculation of the equilibrium $r_s$ values
for closed-shell clusters show similar behaviors as their bulk metals i.e.,
they also show increasing behaviors as functions of $\zeta$. But, the situation
is somewhat different for open-shell clusters. The open-shell clusters show
decreasing behavior at lower polarizations and increasing behaviors at higher
polarizations. The equilibrium $r_s$ values of the ground state configuration
of the clusters are always smaller than the bulk value. This self-compression
is due to the surface tension. On the other hand, at higher polarizations, the
equilibrium $r_s$ values exceeds the bulk value $\bar r_s(0)$, and this is
called self-expansion. In conclusion, the SJM, has provided a method which can
be used to calculate the sizes of the simple metal clusters with minimum
possible efforts. More realistic results for open-shell clusters are possible
when the
spherical geometry for the jellium background is replaced by the spheroidal or
ellipsoidal shapes. Work in this direction is in progress.

%%%%%%%%%%%%%%%%%%%%%%%%%%%%%%%%%%%%%%%%%%%%%%%%%%%%%%%%%%%
\newpage

%%%%%%%%%%%%%%%%%%%%%%%%%%%%%%%%%%%%%%%%%%%%%%%%%%%%%%%%%


\begin{thebibliography}{99}

\bibitem{ekardt}
W. E. Ekardt, Phys. Rev. B {\bf 29}, 1558 (1984).
\label {ekardt}

\bibitem{knight}
W. D. Knight, K. Clemenger, W. A. de Heer, W. A. Saunders, M. Y. Chou, and M.
L. Cohen, Phys. Rev. Lett. {\bf 52}, 2141 (1984).
\label {knight}

\bibitem{brack93}
M. Brack, Rev. Mod. Phys. {\bf 65}, 677 (1993), and references therein.
\label {brack93}

\bibitem{rubio}
A. Rubio, L. C. Balb\'{a}s, and J. A. Alonso, Z. Phys. D{\bf 19}, 93 (1991).
\label{rubio}

\bibitem{lang70}
N. D. Lang and W. Kohn, Phys. Rev. B {\bf 1}, 4555 (1970).
\label {lang70}

\bibitem{ashlang67}
N. W. Ashcroft and D. C. Langreth, Phys. Rev. {\bf 155}, 682 (1967).
\label {ashlang67}

\bibitem{shore99}
H. B. Shore and J. H. Rose, Phys. Rev. B {\bf 59}, 10485 (1999) and references
therein.
\label{shore99}

\bibitem{pertran}
J. P. Perdew, H. Q. Tran, and E. D. Smith, Phys. Rev. B {\bf 42}, 11627 (1990).
\label {pertran}

\bibitem{kohnsham}
W. Kohn and L. J. Sham, Phys. Rev. {\bf 140}, A1133 (1965).
\label{kohnsham}

\bibitem{pay99}
M. Payami, J. Chem. Phys. {\bf 111}, 8344 (1999).
\label{pay99}

\bibitem{perdew93}
J. P. Perdew, M. Brajczewska, and C. Fiolhais, Solid State Commun. {\bf 88},
795 (1993).
\label{perdew93}

\bibitem{braj96}
M. Brajczewska, A. Vieira, C. Fiolhais, and J. P. Perdew, Prog. Surf. Sci.
{\bf 53}, 305\\ (1996).
\label{braj96}

\bibitem{vieira97}
A. Vieira, M. B. Torres, C. Fiolhais, and L. C. Balb\'as, J. Phys. B: At. Mol.
Opt. Phys. {\bf 30}, 3583 (1997).
\label{vieira97}

\bibitem{pay98}
M. Payami and N. Nafari, J. Chem. Phys. {\bf 109}, 5730 (1998).
\label{pay98}

\bibitem{perwan}
J. P. Perdew and Y. Wang, Phys. Rev. B {\bf 45}, 13244 (1992).
\label{perwan}

\bibitem{ash66}
N. W. Ashcroft, Phys. Lett. {\bf 23}, 48 (1966).
\label {ash66}

\end{thebibliography}
\end{document}